\documentclass[aps,prc,12pt]{revtex4-1}
\usepackage{epsfig}

\begin{document}
\title{Comments on Proton Emission after Muon Capture
\footnote{Technical Report - meco-34}}

\author{Ed V. Hungerford}
\affiliation{Department of Physics\\
University of Houston\\
Houston,  TX  77025\\
hunger\@ UH.EDU\\
(713)743-3549\\
FAX:(713)743-3549}

\date{May 13, 1999}

\begin{abstract}

The emission of particles after atomic muon 
capture was studied extensively during
the early 70's.  In particular, proton and neutron
spectra were measured
in order to determine whether nuclear emission was due to a
direct or a thermalized
process.  It was found that protons emitted after atomic muon
capture decreased from a probability of 15\% in light nuclei (C, N, O)
to essentially 0\% for heavy systems such as $^{120}$Sn, and that only 10\%
of this emission could be ascribed to statistical processes.  On the other
hand a significant thermal spectrum of neutrons in addition to a direct
component was found. Unfortunately these studies were not always self
consistent so that information of relevance to MECO is ambiguous.
\end{abstract}

\maketitle

\section{Previous Experimental Results}

There are several summaries of experimental
results\cite{uberall,singer,mukhop} which discuss
the observation and theoretical interpretation
of the emission spectrum of protons, deuterons, and alphas after atomic muon
capture.  Unfortunately the literature is not always self consistent.
It appears that charged particle emission can be as large as 15\% in
Carbon and Oxygen nuclei, decreasing to a few \% in emulsion nuclei
(Ag/Br) and to essentially 0\% in heavier systems such as $^{120}$Sn.
Just how the spectrum divides into the various particle types is not
known, although one study indicates that the emission of alphas
appears to be more consistent with a thermalized distribution than
protons.  The pyroton spectrum is found to have a high energy tail, and is
more consistent with the direct process of muon 
capture on a quasi-deuteron.

\begin{figure}[t] 
  \begin{center}
  \mbox{\epsfig{figure=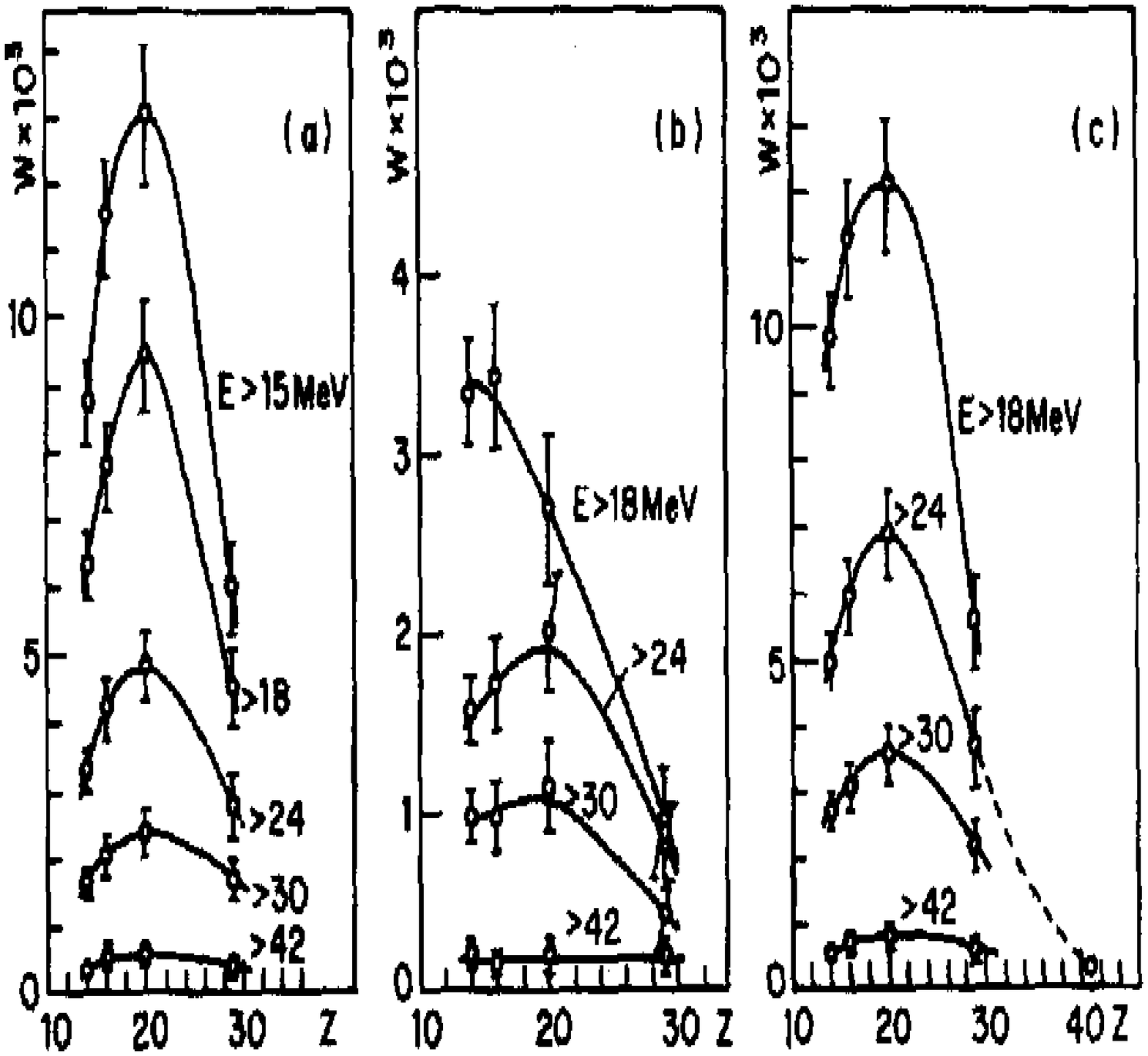, width=4in}}
  \caption{The Spectrum Integral for Charged Particle Emission after
  Muon Capture above Various Energy Thresholds as a Function of Z.
  from Ref.[4]}
   \label{figure1}
   \end{center}
\end{figure}

High energy proton, deuterium, and tritium emission\cite{budyas} 
from $^{28}$Si, $^{32}$S, $^{40}$Ca, and $^{64}$Cu 
is summarized in Figure \ref{figure1}.  In this
figure the partial spectrum integral is given for protons and deuterons.
Here W is the
probability of particle emission per muon capture.  These
data were taken by counters so that the actual spectra have a
low energy cut-off, about 14 MeV for protons.  Thus one sees
that the partial spectrum integral, reflecting
high energy proton emission, peaks at Z=20.  Unfortunately the 
proton spectrum
below 14 MeV is not known, and the contribution from lower 
energy emissions to this integral is undertermined.

\begin{figure}[t] \begin{center}
  \mbox{\epsfig{figure=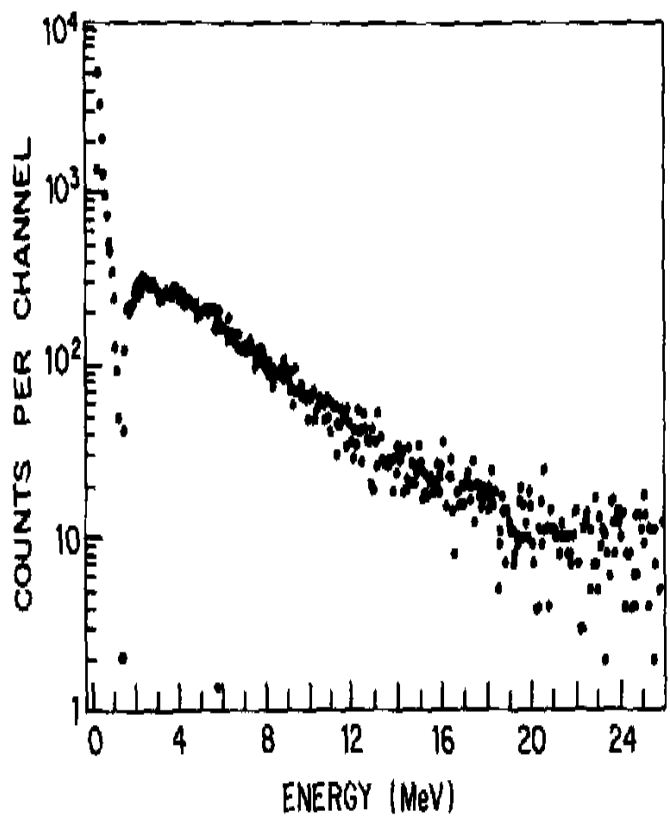,width=4in}}
  \caption{The Charged Particle Emission Spectrum from Si Corrected
  for Muon Decay Electrons.  from Ref.[5]}
  \label{figure2}
  \end{center}
\end{figure}

The total charged particle spectrum has also been measured in a Si(Li) 
detector\cite{sobottka}, and is shown in Figure \ref{figure2}.  
This spectrum is
corrected for electrons from muon decays and proton emissions from this
detector.  Obviously it has a high-energy cut-off at 25 MeV.  The 
charged particle spectrum from the Si target obtained in Ref.[4] matches
these data in the region of overlap. The spectrum shape shows
a continual exponential
decrease in differential emission probability from below the Couloumb
barrier, out to some 50 MeV.  The
low energy rise in the spectrum below the break at about 1.4 MeV is
associated with $^{27}$Al recoils, and is consistent with thermal neutron
emission in the reaction;

\vspace{0.2cm}
\noindent \hspace*{2cm}$\mu + ^{28}\mbox{Si}\rightarrow ^{27}\mbox{Al}
+ n + \nu_{\mu}$.\\

\noindent The exponential decay constant of the spectrum tail
is about 3.1 MeV, and the spectrum peaks at 2.5 MeV, somewhat below
the Couloumb barrier of 4.6 MeV.  The total spectrum
integral gives a probability for charged particle emission of 15\% per
$\mu$ capture.  Note that the spectrum integral from the Si 
data of Ref.[4] for
energies above 18 MeV is only about 1\%, indicating that most
of the charged particles emitted from 
this nucleus have energies less than 18 MeV.
This reflects an experimental problem when attempting to observe 
charged particle emission with counters external to the target.

Several experiments used radiochemical techniques to determine the
probability of charged particle emission.  This can give
the full spectrum integral for particle emission if a unique reaction
leaving a radioactive daughter can be identified.  
An example is shown in \ref{figure2}, where the charged
particle yield per stopped muon for several nuclei is shown for the
reactions $(\mu;\mbox{p,2n}), (\mu;\mbox{pn}), \mbox{and
}(\mu;\mbox{p})$.  One can compare the sum of the yield from these
reactions for Mn, about 2\%, to the data of Ref.[4], also about 2\%
for the sum of proton and deuteron emission, but for particle 
energies above 15
MeV.  To compare these yields, the data may both be normalized 
to the yield per $\mu$-capture.
Assuming that the counter data is correct, it 
appears that the radiochemical yield must
be low unless the spectrum for Mn below 15 MeV falls very rapidly to
zero, unlike the data for Si.

There are also radiochemical measurements
for Si \cite{vil}, however these data only observe the $(\mu;\mbox{p})$
reaction.  They indicate that 5\% of the $\mu$-captures lead to proton
emission.  If these data and those of Ref.[5] are correct, then there
must be significant, simultaneous neutron and proton emission.  Indeed
this does appear to be the case in the radiochemical data of Ref.[4],
but these data were for heavier targets.  Still the
radiochemical data seems to give much too low a yield.

\begin{figure}[ht] \begin{center}
  \mbox{\epsfig{figure=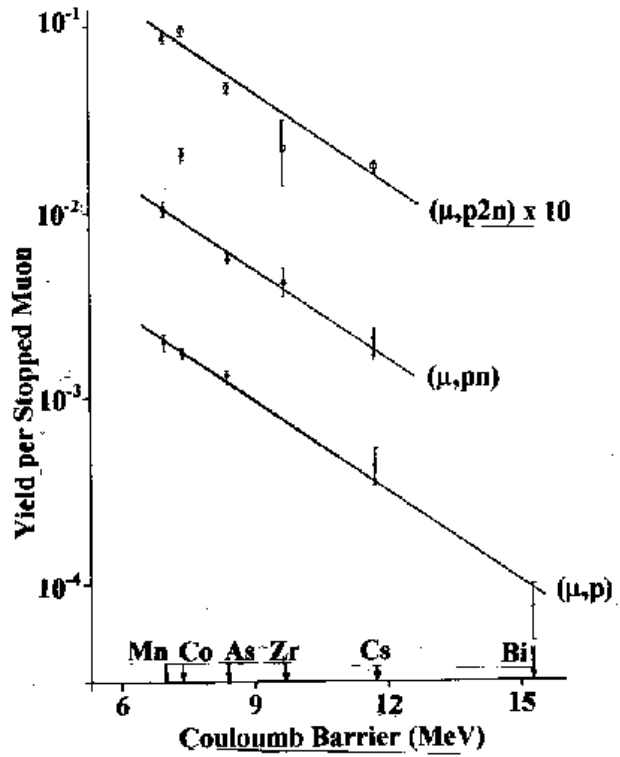,height=4in}}
  \caption{Radiochemical Analysis of the Charged particle Yield after
  Muon Capture as a Function of the Couloumb Barrier.  from Ref.[6]}
   \label{figure3}
   \end{center}
\end{figure}
In summary, it appears despite some uncertainties, that charged
particle emission from light nuclei can be as high as 15\%
(consistent in several experiments). The energy spectrum
rises to a peak near the Couloumb barrier (1-4MeV) and then decays
exponentially up to high energies, $\ge$ 50 MeV.  Proton emission with
simultaneous neutron emission predominates, but other charged
particles are also emitted in measurable quantities.

\section{Impact on MECO}

Meco captures stopping muons in an $^{27}$Al target.  The 
signal of interest is due to an 
electron, emitted with approximately 100 MeV/c momentum.  Thus if
the detector is tuned for this momentum range, protons of kinetic
energy above 5.3 MeV may cause significant background.  I
assume below that the Si spectrum of Figure \ref{figure3}
represents the charged particle
spectrum from $^{27}$Al, and normalize the 
spectrum integral to obtain
15\% charged particle emission after a $\mu$ capture.  If one then assumes
that the proton/deuteron ratio is the same for this integral as was
measured in Ref.[4] for energies above 18 MeV (probably not a good assumption), 
then about 65\% of this spectrum are
protons and 35\% deuterons.  One notes that deuterons,
may also create substantial background, but because of their higher
magnetic rigidity, the field of the detector selects
those particles from a higher energy window.
\begin{figure}[ht] \begin{center}
  \mbox{\epsfig{figure=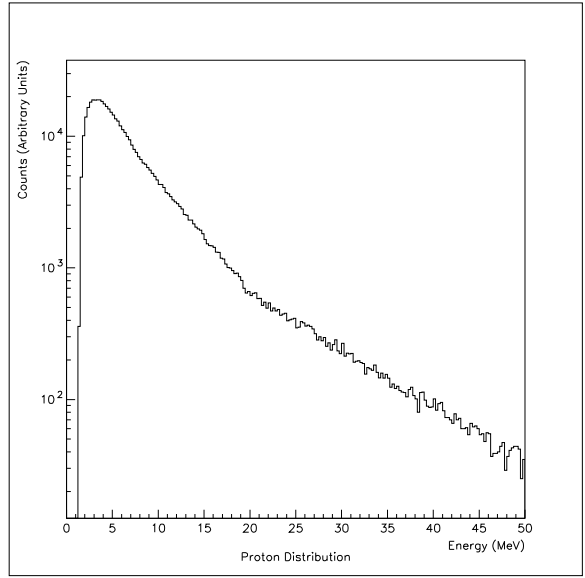,height=3in}}
  \caption{A Plot of the Funtion Fitted to the Proton Emission
  Spectrum from Si}
  \label{figure4}
   \end{center}
\end{figure}

Previously the proton spectrum I generated for Meco was 
obtained from a spectrum shape fit
to the exponential proton tail and normalized by the radiochemical
experiments.  It now appears that this must underestimate the
number of low energy protons.  Thus
I now fit the Si spectrum of Ref.[5] with a function of the form;

\begin{eqnarray}
\noindent\hspace*{2cm}\mbox{W(T)} &=&A_{1}
(1-\mbox{T}_{th}/\mbox{T})^{\alpha}
e^{-\mbox{T/T}_{1}} \qquad \mbox{E} \geq 8 \mbox{MeV}; \\
			   &=& A_{2}e^{-\mbox{T/T}_{2}} \qquad 
8 \mbox{MeV} \leq \mbox{T} \leq 20 \mbox{MeV}; \\
                           &=& A_{3}e^{-\mbox{T/T}_{3}}\qquad 
\mbox{T} \geq 20 \mbox{MeV}.\\
\end{eqnarray}

\vspace{0.2cm}
\noindent where T is the kinetic energy and the fitted parameters 
are; A$_{i}= 0.092405, 5.4374^{-3}, 5.1785^{-4}$ MeV$^{-1}$, 
T$_{th}$=1.4 Mev, $\alpha$=1.3279 , 
and T$_{i}$ = 3.1, 5.1, 10.0 MeV.
The shape of this spectrum is shown in Figure \ref{figure4}.  
It has a low energy cut-off
at 1.4 Mev, just below the Couloumb barrier at 4.6 MeV.

\section{Conclusions}

A more careful study of the number of charged particles emitted after
atomic muon capture indicates that previous estimates of this spectrum
were underestimated.  Therefore a new spectrum generator was created
which better represents the measured charged spectrum from Si.  This
spectrum is normalized to other data which indicate that the yield of
protons is 10\%, and the yield of deutrons is 5\%
per stopped muon.

Finally, one notes that a target heavier than Al could significantly
decrease the number of emitted charged particles, although the time
for muon capture in a heavier nucleus decreases, reducing the
sensitivity to muon conversion.

\end{document}